\documentclass[twocolumn,prl,showpacs,preprintnumbers,amsmath,amssymb,floatfix]{revtex4}

\usepackage{graphicx}
\usepackage{booktabs}
\usepackage{float}

\begin{document}

\title{Transition-Metal Substitutions in Iron Chalcogenides}

\author{V. L. Bezusyy, D. J. Gawryluk, A. Malinowski, and Marta Z. Cieplak}

\affiliation{Institute of Physics, Polish Academy of Sciences, Al. Lotnik\'{o}w 32/46, 02-668 Warsaw, Poland}

\begin{abstract}

The $ab$-plane resistivity and Hall effect are studied in Fe$_{1-y}$M$_y$Te$_{0.65}$Se$_{0.35}$ single crystals doped with two transition metal elements, M = Co or Ni, over a wide doping range, $0 \leq y \leq 0.2$. The superconducting transition temperature, $T_{c}$, reaches zero for Co at $y \simeq 0.14$ and for Ni at $y \simeq 0.032$, while the resistivity at the $T_{c}$ onset increases weakly with Co doping, and strongly with Ni doping. The Hall coefficient $R_H$, positive for $y$ = 0, remains so at high temperatures for all $y$, while it changes sign to negative at low $T$ for $y > 0.135$ (Co) and $y > 0.06$ (Ni). The analysis based on a two band model suggests that at high $T$ residual hole pockets survive the doping, but holes get localized upon the lowering of $T$, so that the effect of the electron doping on the transport becomes evident. The suppression of the $T_c$ by Co impurity is related to electron doping, while in case of the Ni impurity strong electron localization most likely contributes to fast decrease of the $T_c$.

\end{abstract}

\maketitle


Substitution of impurities is an efficient method for tuning of the electronic properties and mapping out the phase diagrams of the new compounds. While isovalent substitutions are expected to be mainly potential scatterers, the heterovalent impurities usually induce doping of carriers. Many recent studies focus on iron-based superconductors (IS), in which pairing mechanism is likely related to spin fluctuations \cite{Johnston2010,Wen2011,Hirschfeld2011,Stewart2011}. Since the IS are multiband materials, the effect of impurities on the phase diagram may be quite complex. For example, the heterovalent substitution of the transition metals into Fe-site in iron pnictide BaFe$_2$As$_2$ induces not only shift of the Fermi level \cite{Canfield2009,Fang2009,Canfield2010,Olariu2011,Li2012} but also reconstructs the Fermi surface \cite{Liu2011}. Both the shift of the chemical potential and loss of the coherent carrier density are predicted by theory \cite{Berlijn2012}.

Less is known about substitutions in iron chalcogenides, another class of the IS. Few attempts of the transition metal doping of single crystals of FeTe$_{1-x}$Se$_x$ have been reported, and for a limited impurity contents \cite{Nabeshima2012,Zhang2013,Inabe2013}. The studies are complicated by two types of magnetic correlations existing throughout the phase diagram. While the $(\pi,\pi)$ correlations are likely involved in the superconducting pairing mechanism, the $(\pi,0)$ correlations, involved in the magnetic ordering in the parent compound Fe$_{1+\delta}$Te, survive on a local scale upon Se doping, and likely contribute to incoherent scattering of carriers \cite{Bao2009,Liu2010,Hu2013}. The incoherent scattering may be further enhanced by inhomogeneities present in the crystals, such as Fe excess \cite{Bao2009,Liu2009,Thampy2012} or Fe vacancies \cite{Wittlin2012,Bezusyy2014}.

In this Letter we report on the first comprehensive study of the transport properties of FeTe$_{0.65}$Se$_{0.35}$ single crystals, doped over a wide doping range by transition metal elements, Co and Ni. While the crystals of FeTe$_{1-x}$Se$_{x}$ with $x = 0.5$ are optimal for superconductivity, two tetragonal phases with slightly different Se content have been reported \cite{Sales2009}. On the other hand, the crystals with $x = 0.35$ are grown as a single-phase material, and, out of 18 elements that have been examined, only Co, Ni and Cu are found to substitute Fe-site \cite{Gawryluk2011}. Our study reveals that the Fermi surface, which in undoped material consists of hole and electron pockets \cite{Chen2010}, evolves dramatically with doping, in a manner resembling the evolution in doped pnictides \cite{Liu2011}. However, the influence of impurities on superconductivity shows features distinct from pnictide superconductors.


Single crystals of Fe$_{1-y}$M$_y$Te$_{0.65}$Se$_{0.35}$ with M = Co or Ni and $y$ up to 0.2 have been grown using Bridgman's method \cite{Gawryluk2011}. The X-Ray Powder Diffraction (XRPD) shows that all samples have P4/nmm tetragonal matrix with small amount of secondary phases, mainly Fe$_3$O$_4$ and hexagonal Fe$_7$(Te-Se)$_8$ \cite{Gawryluk2011,Bezusyy2012}, which are often present in such crystals \cite{Taen2009,Williams2009}. High resolution transmission electron microscopy reveals that the size of Fe$_7$(Te-Se)$_8$ inclusions decreases with increasing velocity of crystallization \cite{Wittlin2012}. To keep the inclusions small, in the present study this velocity has been kept high, above 15 mm/h. The quantitative point analysis, performed by Energy-Dispersive X-ray (EDX) spectroscopy in many different points on each crystal, shows that Co and Ni effectively substitute Fe and their contents are close to the nominal. The average Fe+M content is 0.99(3) and the Se content is 0.35(2). The evaluation of the magnetic properties of the doped crystals, which will be described elsewhere \cite{Puzniak}, shows neither localized magnetic moments, no any magnetic order developing as a result of doping.
The resistivities, in-plane ($\rho$), and Hall ($\rho_{xy}$), were measured by $dc$ and $ac$ four-probe methods, respectively, using Physical Property Measurement System (Quantum Design), in the temperature range 2 to 300 K, and in magnetic fields up to 9 T. The $ac$ magnetic susceptibility was measured with magnetic field amplitude of 1 Oe and a frequency of 10 kHz in warming mode (field orientation has no effect on the superconducting transition temperature, $T_c$).

\begin{figure}
\includegraphics[width=8.5cm]{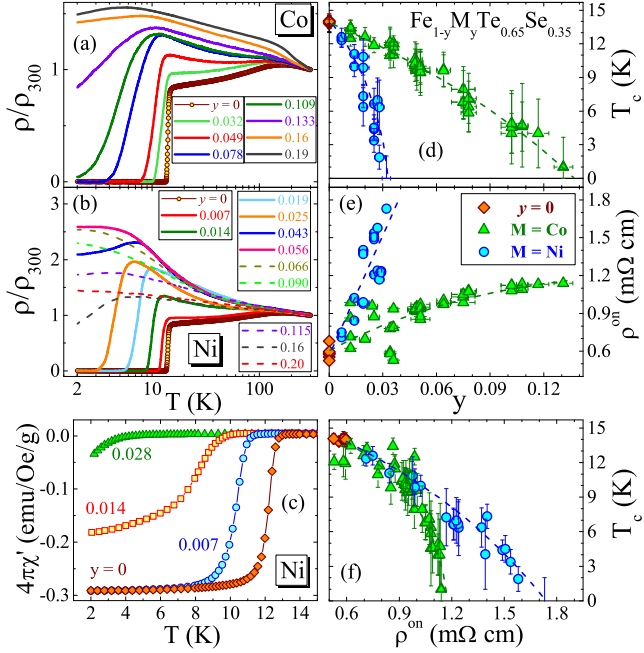}
\caption{(Color online) $\rho/\rho_{300}$ versus $T$ for crystals doped with Co (a) and Ni (b). (c) The $ac$ susceptibility for several Ni-doped crystals (the data are not corrected for demagnetization factor). (d) $T_c$ vs $y$, (e) $\rho^{on}$ vs $y$, and (f) $T_c$ vs $\rho^{on}$, for undoped crystals (brown diamonds), and for Co- (green triangles), and Ni-doped crystals (blue circles). In (d-f) the dashed lines are guides to the eye. }
\label{fig:Rho}
\end{figure}
\begin{figure}
\includegraphics[width=8.5cm]{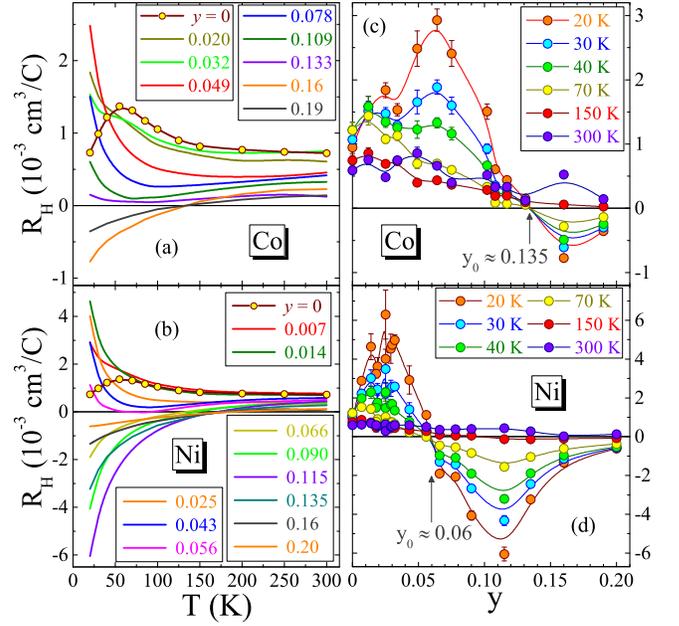}
\caption{(Color online) $R_H$ as a function of $T$ for crystals with different $y$, doped with Co (a) and Ni (c), and as a function of $y$ for various $T$ for Co (c) and Ni (d).}
\label{fig:Hall}
\end{figure}

The $T$-dependence of the $\rho$, normalized to resistivity at room temperature, ${\rho}_{300}$, is shown in Fig.\ref{fig:Rho}(a-b) for crystals with different doping contents, $y$. In the undoped crystal, $y = 0$, the $\rho$/${\rho}_{300}$ decreases with $T$ decreasing below about 150 K, indicating good metallic character. Fig.\ref{fig:Hall}(a-b)shows the $T$-dependence of the Hall coefficient, $R_H$, for the same crystals, in the high-$T$ range above 20 K, in which the dependence of the Hall resistance on the magnetic field is linear. At $y = 0$ the $R_H$ is positive; it rises as $T$ is lowered down from 300 K to about 55 K, and has a downturn at lower $T$. Such a behavior, reported before \cite{Liu2009}, results from the interplay between different $T$-dependencies of charge carrier densities and their mobilities in the multiband system. The $T$-dependencies of $\rho /\rho_{300}$ for doped crystals acquire low-temperature upturns, which increase progressively with increasing $y$. The upturns are much larger in case of Ni doping. They appear also in the $R_H(T)$ dependencies at small $y$ \cite{upturns}. Unexpectedly, in case of Ni-doped crystals with $y > 0.06$, the $\rho /\rho_{300}$ at low $T$ shows nonmonotonic changes with increasing $y$: first saturation, than decrease, and finally increase at the highest doping [Fig.\ref{fig:Rho}(b)].

The susceptibility data, shown in Fig.\ref{fig:Rho}(c) for several Ni-doped crystals, confirm bulk superconductivity for $y < 0.03$. Fig.\ref{fig:Rho}(d) displays the $T_c$ versus $y$ for both impurities. Here $T_{c}$ is defined as the middle point of the transition, and the vertical errorbars reflect 90\% to 10\% transition width. The $y$ values and the horizontal errorbars are the average impurity content and the standard deviations, respectively, obtained from several EDX data measured in different points on each sample. The $T_c$ is about 14 K at $y = 0$. It is suppressed by both impurities, reaching zero at critical concentrations $y_c \simeq 0.14$ (Co) and $y_c \simeq 0.032$ (Ni). The initial slopes of the $T_c (y)$ dependencies, $dT_c/dy$, are equal to about -0.9 and -3.5 K/at.\% for Co and Ni, respectively. Assuming that superconductivity is suppressed by a rigid band shift due to electron doping, one expects roughly twice as fast decrease of the $T_c$ to zero for the Ni than for the Co, because Ni provides twice as many electrons to the electronic bands. This, however, is not the case. The Ni-induced suppression is much more effective. Thus, it cannot be explained by electron doping alone. It is likely that deeper impurity potential results in more complex modification of the band structure, or strong localization of carriers.

The nonlinear $T$-dependence of the resistivity prevents easy estimate of scattering rates. Instead, we extract from the data normalized $\rho$ at the onset of superconductivity, $\rho_{on} /\rho_{300}$. To eliminate the scatter of the data due to imperfect estimate of the sample dimensions, we make use of the fact that $\rho_{300}$ does not show any definite dependence on $y$, so we may average it over various $y$ to obtain $\overline{\rho_{300}}$. Finally, we define the quantity, $\rho^{on} \equiv (\rho_{on}/ \rho_{300})\overline{\rho_{300}} \simeq \rho_{on}$, which gives good approximate value of the resistivity at the $T_c$ onset. Fig.\ref{fig:Rho}(e) shows the $\rho^{on}(y)$ dependence for both impurities. We find that the initial increase of $\rho^{on}$ with increasing $y$ is by a factor of 4.5 times larger in the Ni case, indicating much stronger scattering induced by the Ni. Fig.\ref{fig:Rho}(f) compares the dependencies of the $T_c$ on $\rho^{on}$ for both impurities. They are distinctly different. At the point at which the $T_c$ reaches zero the increase of $\rho^{on}$ induced by the Ni is about twice as big as the increase induced by the Co. 

Turning the attention to $R_H$ [Fig.\ref{fig:Hall}(a-b)], we observe that with increasing $y$ the magnitude of $R_H$ is reduced in the whole temperature range, what is consistent with the expectation of electron doping. Eventually, above certain impurity content $y_0$, and below certain temperature, $T_0$, the $R_H$ changes sign into negative. The $R_H(y)$-dependencies for various $T$ are collected in Fig.\ref{fig:Hall}(c-d) for Co- and Ni-doped crystals, respectively. It is seen that at 300 K the $R_H$ depends weakly on $y$ and remains positive for all $y$-values, but significant dependencies develop upon cooling. Qualitatively similar behavior is observed for both impurities, with two broad features developing upon lowering of $T$, the enhanced positive $R_H$ at $y < y_0$, and the enhanced negative $R_H$ at $y > y_0$. Both features are stronger in the case of Ni impurity. The value of $y_0$ at low $T$ in Ni-doped crystals ($y_0 \simeq 0.06$) is about twice as small as in the case of Co ($y_0 \simeq 0.135$), what confirms that electron doping contributes to the behavior. Note also that the low-$T$ decrease of $\rho$ in Ni-doped crystals occurs for $y \gtrsim y_0$, suggesting relation to electron doping.

\begin{figure}
\includegraphics[width=8.5cm]{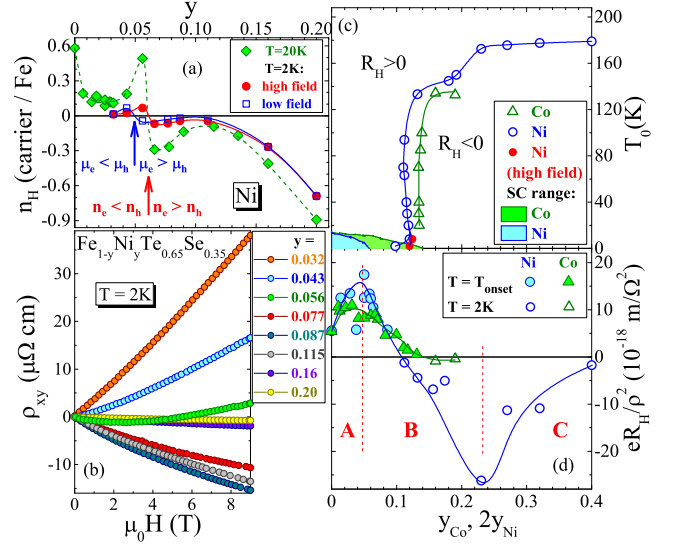}
\caption{(Color online) (a) $n_H$ versus $y$ for Ni impurity at $T = 20 K$ (full green diamonds), and at $T = 2$ K: high field (full red circles) and low field (open blue squares). Here, $n_H$(carriers/Fe) = $0.88 \times 10^{-22} n_H$(cm$^{-3}$). (b) $\rho_{xy}$ versus $\mu_0 H$ at $T = 2$ K for Ni-doped crystals. (c) $T_0$ versus $y$ for Co (triangles), and Ni (circles); Ni high-field data: full red circles. Shaded areas: superconductivity (SC) range. (d) $eR_H / \rho^2$ at the $T_c$ onset (full points) and at $T = 2$ K (open points) versus $y_{Co}$ and $2y_{Ni}$ for Co- and Ni-doped crystals, respectively. All lines are guides to the eye.}
\label{fig:nH}
\end{figure}

In Fig.\ref{fig:nH}(a) we plot the $y$-dependence of the Hall number, $n_H = 1/eR_H$, for Ni-doped crystals at $T = 20$ K. The $n_H$ is positive, hole-dominated at $y < y_0$, and becomes negative, electron-dominated at $y > y_0$, with a change of sign at $y_0 = 0.06$. Large magnitude of negative $n_H$ at $y = 0.2$ suggests that at the largest $y$ and at low $T$ the hole contribution becomes small. At all other $y$ both hole and electron contributions are important.

More insights are provided by the dependence of the Hall resistivity, $\rho_{xy}$, on the magnetic field, $\mu_0 H$, measured at $T = 2$ K for all Ni-doped crystals, which do not superconduct down to $T = 2$ K. As shown in Fig.\ref{fig:nH}(b), the dependence evolves, from almost linear with positive slope for $y = 0.032$ ($y < y_0$), to linear with negative slope for $y = 0.2$ ($y > y_0$); it is nonlinear for all intermediate $y$. In case of the crystal with $y =0.056$ ($y \simeq y_0$) we see the negative slope at low $H$, and positive slope at large $H$. In a two band model in the strong-field limit the $R_H$ is dependent on the $\mu_0 H$ \cite{Hall},

\begin{eqnarray}
eR_H = \frac{\mu_{h}^{2} n_{h} - \mu_{e}^{2} n_{e} + \mu_{e}^{2}\mu_{h}^{2}(n_h - n_e) (\mu_0 H)^2}{\rho^{-2} + \mu_{e}^{2}\mu_{h}^{2}(n_h - n_e)^2 (\mu_0 H)^2} \nonumber \\
\overrightarrow{_{H \rightarrow \infty}} \quad \frac{1}{n_h -n_e},
\label{strongH}
\end{eqnarray}

while in the low-field limit it is given by $eR_H = \rho^2 (\mu_{h}^{2} n_{h} - \mu_{e}^{2} n_{e})$. Here $n_h$ and $n_e$ are hole and electron concentrations, $\mu_h$ and $\mu_e$ are their mobilities, and $\rho = (\mu_{h} n_{h} + \mu_{e} n_{e})^{-1}$ is the resistivity at $H = 0$. From the 2K-data we extract the $n_H$ in the low-field, and in the high-field limits, and we show it as a function of $y$ in Fig.\ref{fig:nH}(a). The high-field $n_H$ changes sign at $y \simeq 0.06$, indicating, according to Eq.(\ref{strongH}), that $n_e > n_h$ for $y > 0.06$. On the other hand, the low-field $n_H$ changes sign at $y \simeq 0.05$, when $n_h$ is still larger than $n_e$; therefore, we have $\mu_e > \mu_h$ for $y > 0.05$. Arrows indicate in the figure the $y$ values, at which these changes occur.

We summarize the Hall effect data by plotting in Fig.\ref{fig:nH}(c) the $y$-dependence of the temperature $T_0$, at which the $R_H$ changes sign. The Ni content is multiplied by two, to allow for comparison of the electron-doping by the two impurities. It is seen that the region with $R_H <0$ is restricted to $T < T_0$, and to $y > y_0$. At largest doping, $2y_{Ni} \gtrsim 0.23$, the boundary between positive and negative $R_H$ is located at $T_0 \approx 180$ K. The observation of hole-dominated conduction at high temperature in heavily doped crystals indicates that hole pockets survive the electron doping. However, these hole pockets must be substantially shrank in comparison with the undoped crystal. The shrinking of the hole pockets has been observed by angle-resolved photoemission (ARPES) in Co-doped pnictide BaFe$_2$As$_2$ \cite{Liu2011}. One may expect that the hole carriers in such remnant hole pockets are easily localized when temperature is lowered, what explains the electron-dominated conduction below $T_0$. Interestingly, at slightly smaller doping, $0.12 < 2y_{Ni} < 0.23$, the $T_0$ is reduced to form a plateau at about 140 K. Similar plateau at similar value of $T_0$ is seen in Co-doped samples in the doping range $0.14 < y_{Co} < 0.2$, suggesting common origin of this feature. It may signal some reconstruction of the Fermi surface, such as, for example, vanishing of one of the hole pockets, which have been observed by ARPES at the $\Gamma$ point in the undoped crystal \cite{Chen2010}.

Combining the data for $R_H$ and $\rho$, we can obtain in the weak field limit the quantity $eR_H/\rho^2 = \mu_{h}^{2} n_{h} - \mu_{e}^{2} n_{e}$. Fig.\ref{fig:nH}(d) shows the plot of $eR_H/\rho^2$ versus $y_{Co}$ and $2y_{Ni}$. The low-$T$ data are calculated for all Co- and Ni-doped crystals, either at the onset of superconductivity (when it is present), or at $T = 2$ K in case of nonsuperconducting samples. We see three regions with distinct dependencies of $eR_H/\rho^2$ on $y$. At small $y$ (labeled \textbf{A}) the $eR_H/\rho^2$ increases with increasing $y$, and the increase is larger for Ni impurity. At intermediate doping (region \textbf{B}) there is a profound decrease of $eR_H/\rho^2$ until it reaches negative value, particularly large in case of the Ni. Finally, at the largest Ni doping (region \textbf{C}) we observe the decrease of the magnitude of negative $eR_H/\rho^2$.

We now consider the effect of electron doping on carrier concentrations and mobilities, assuming for simplicity that only monotonic changes occur. We anticipate that $n_h$ decreases and $n_e$ increases with increasing $y$; also, the $\mu_h$ is expected to decrease due to impurity scattering and shrinking of the hole pockets. On the other hand, the $\mu_e$ is influenced by two competing effects, impurity scattering, which reduces $\mu_e$, and the expansion of the electron pockets, which is likely to increase $\mu_e$. If the last effect prevails, the $eR_H/\rho^2$ should be a decreasing function of $y$. This expectation agrees nicely with the observation of the profound decrease of $eR_H/\rho^2$ in the region \textbf{B}. When the increasing $\mu_e^2 n_e$ exceeds decreasing $\mu_h^2 n_h$, the $R_H$ changes sign; furthermore, this evolution explains the decrease the $\rho$ in the low-$T$ limit illustrated in Fig.\ref{fig:Rho}(b).

In the region \textbf{C} the hole contribution is substantially reduced, so that $\mu_{e}^{2} n_{e}$ becomes the dominant term. Scattering by large density of impurities is likely to reduce the $\mu_{e}^2$, what overweighs the increase of $n_e$, leading to the decrease of the magnitude of $eR_H/\rho^2$.

Similarly, in this simple picture the initial increase of $eR_H/\rho^2$ in the region \textbf{A}, larger for Ni impurity, may be explained only by strong impurity-induced reduction of $\mu_e^2$, which overweighs the minor increase of $n_e$ at small doping; moreover, the decrease of $\mu_{e}^2 n_e$ would have to be larger than the decrease of $\mu_{h}^2 n_h$. A conceivable scenario could be the low-$T$ localization of electron carriers in the vicinity of positively charged impurity ions, which should be more profound in case of the Ni with larger impurity potential. Such scenario is supported by the observation of strong increase of $\rho^{on}$ in Ni-doped crystals. Eventually, with the further increase of $y$ into the \textbf{B} region the expansion of the electron pockets takes over, and the electron localization cease to affect transport. We should caution here that this reasoning is based on a simple two band model, which may not be entirely appropriate for this multiband system. However, it seems to provide good qualitative explanation of the observed behavior.

The complicated evolution of the multiband system, uncovered in this work, prevents estimating the scattering rates of different types of carriers on the basis of present experiment, since this requires independent evaluation of carrier concentrations and mobilities. Therefore, we cannot compare the observed rate of the $T_c$ suppression to the theoretical predictions of pair-breaking for various gap symmetries, similar to what has been done in case of doped or irradiated pnictides \cite{Li2012,Prozorov2014}. Nevertheless, our results suggest that the electron localization at small doping levels may play a role in the destruction of superconductivity. As indicated in Fig.\ref{fig:nH}(c), where superconducting regions are shown by shaded areas, in Co-doped crystals the $y_c$, at which $T_c$ reaches zero, is comparable to $y_0$. Therefore, it is likely that in this case the suppression of the $T_c$ is related to the shrinking of hole pockets, similar to the effect identified by ARPES in Co-doped BaFe$_2$As$_2$ \cite{Liu2011}. On the other hand, in case of Ni-doped crystals the $y_c$ is much smaller than $y_0$, in fact, it is close to the boundary of the \textbf{A} region. This suggests that electron localization, particularly strong in the Ni case, may contribute, in addition to electron doping, to faster suppression of the $T_c$. This result is different from the behavior reported for Ni-doped BaFe$_2$As$_2$, for which Ni impurity appears to be much stronger electron scatterer than the Co (as in the present case), but the $T_c$ suppression seems to be well explained solely by electron doping \cite{Canfield2010,Olariu2011}. It is intriguing to ask what is the origin of such a strong difference between these two materials. It is possible that the answer lies in the persistence of local $(\pi,0)$ magnetic fluctuations in FeTe$_{0.65}$Se$_{0.35}$ \cite{Hu2013}, which may enhance incoherent scattering in the presence of deep Ni-impurity potential. The other possibilities include local lattice distortion around impurity, which may differ depending on the host lattice, or other subtle difference in the evolution of the electronic structure with doping.

In conclusion, we have studied for the first time the transport properties of FeTe$_{0.65}$Se$_{0.35}$ single crystals doped up to high impurity levels with two transition-metal elements, Co and Ni. At low temperatures the $R_H$ changes sign to negative for $y > 0.135$ (Co) and $y > 0.06$ (Ni), consistent with the electron doping induced by both impurities. However, the $R_H$ remains positive at high $T$, suggesting that remnant hole pockets survive the doping, and holes get localized upon the lowering of $T$. The $T_c$ decreases to zero at $y_c \simeq 0.14$ (Co), and 0.032 (Ni), while the resistivity at the $T_c$ onset is weakly affected by the Co, and it increases strongly for the Ni. These results suggest that the suppression of the $T_c$ is related to electron doping in case of Co impurity, while Ni impurity most likely induces, in addition, strong electron localization.


We would like to thank M. Berkowski, M. Koz{\l}owski, and A. Wi\'{s}niewski for experimental support and discussions. This research was partially supported by the ERC FunDMS Advanced Grant (FP7 Ideas) and by the Polish NCS grant 2011/01/B/ST3/00462; and was partially performed in the NanoFun laboratories co-financed by the ERDF Project POIG.02.02.00-00-025/09.

\end{document}